\documentclass[twocolumn,showpacs,aps,citeautoscript,superscriptaddress]{revtex4-1}
\usepackage{siunitx}
\usepackage{amsmath}
\usepackage{amsfonts}
\usepackage{color}
\usepackage{amssymb}
\usepackage{textgreek}
\usepackage{subfig}
\usepackage{multirow}
\usepackage{braket}
\usepackage{breqn}
\usepackage{tabularx}
\usepackage{booktabs}
\usepackage{graphicx}
\usepackage{ragged2e}
\usepackage{placeins}
\usepackage{hhline}
\usepackage{float}
\usepackage{dcolumn}
\usepackage{bm}
\usepackage{mhchem}
\usepackage{numprint}
\usepackage{lipsum}
\usepackage[table]{xcolor}
\usepackage[para,online,flushleft]{threeparttable}
\usepackage{hyperref}
\usepackage{rotating}
\AtBeginDocument{
  \heavyrulewidth=.08em
  \lightrulewidth=.05em
  \cmidrulewidth=.03em
  \belowrulesep=.65ex
  \belowbottomsep=0pt
  \aboverulesep=.4ex
  \abovetopsep=0pt
  \cmidrulesep=\doublerulesep
  \cmidrulekern=.5em
  \defaultaddspace=.5em}

\begingroup\makeatletter
\catcode`,=\active
\global\let\breqn@comma,
\protected\gdef,{\ifmmode\expandafter\breqn@comma\else\expandafter\active@comma\fi}
\endgroup

\begin{document}
\title{Quantum electrodynamic corrections for molecules: 
Vacuum polarisation and electron self energy in a two-component 
relativistic framework}
\thanks{Parts of this work were reported in preliminary form in K. Janke,
M.Sc. thesis, Philipps-Universit{\"a}t Marburg, 2023, and in A. E. Wedenig,
B.Sc. thesis, Philipps-Universit{\"a}t Marburg, 2022.}

\author{Kjell Janke}
\affiliation{Fachbereich Chemie, Philipps--Universit{\"a}t Marburg,
Hans-Meerwein-Stra{\ss}e 4, 35032 Marburg, Germany}
\author{Andr\'es Emilio Wedenig}
\affiliation{Fachbereich Chemie, Philipps--Universit{\"a}t Marburg,
Hans-Meerwein-Stra{\ss}e 4, 35032 Marburg, Germany}
\author{Peter Schwerdtfeger} 
\affiliation{Centre for Theoretical Chemistry and Physics, The New Zealand
Institute for Advanced Study (NZIAS), Massey University Albany, Private Bag
102904, Auckland 0745, New Zealand}
\affiliation{Laboratoire Kastler Brossel, Sorbonne Universit{\'e}, CNRS, 
ENS-PSL Research University, Coll{\'e}ge de France}
\author{Konstantin Gaul}
\affiliation{Fachbereich Chemie, Philipps--Universit{\"a}t Marburg,
Hans-Meerwein-Stra{\ss}e 4, 35032 Marburg, Germany}
\author{Robert Berger}
\affiliation{Fachbereich Chemie, Philipps--Universit{\"a}t Marburg,
Hans-Meerwein-Stra{\ss}e 4, 35032 Marburg, Germany}

\date{\today}

\begin{abstract}
\noindent Vacuum polarisation (VP) and electron self energy (SE) are 
implemented and evaluated as quantum electrodynamic (QED) corrections in a 
(quasi-relativistic) two-component zeroth order regular approximation (ZORA) framework. 
For VP, the Uehling potential is considered, and for SE, the 
effective potentials proposed by Flambaum and Ginges as well as the one 
proposed by Pyykkö and Zhao.
QED contributions to ionisation energies of various atoms and 
group 2 monofluorides, group 1 and 11 valence orbital energies, 
$^2\mathrm{P}_{1/2} \leftarrow {}^{2}\mathrm{S}_{1/2}$ and $^{2}\mathrm{P}_{3/2} \leftarrow {}^{2}\mathrm{S}_{1/2}$ 
transition energies of Li-, Na-, and Cu-like ions of nuclear charge 
$Z$ = 10, 20, ..., 90 
as well as $\Pi_{1/2}\leftarrow \Sigma_{1/2}$ and $\Pi_{3/2}\leftarrow\Sigma_{1/2}$ 
transition energies of BaF and RaF 
are presented.
Furthermore, perturbative and self-consistent treatments of QED
corrections are compared for Kohn--Sham orbital energies of gold.
It is demonstrated, that QED corrections can be obtained in a two-component 
ZORA framework efficiently and in excellent agreement with corresponding
four-component results.
\end{abstract}

\maketitle

\section{Introduction}
Relativistic effects play a crucial role for the quantitative description of
heavy-element containing molecules \cite{pyykko:1988, pyykko:2012b}. The usual starting point for a relativistic
description of the electronic motion in a molecule is the free-particle Dirac
equation \cite{dirac:1928a, dirac:1928b} combined with a classical Coulomb-type potential for the description
of leading order particle-particle interactions \cite{saue:2011}. This formalism describes the
motion of electrons in accord with special relativity, but not its
interactions, which are due to the Coulomb potential instantaneous. Within
quantum electrodynamics (QED), i.e. the relativistic theory of the
electromagnetic force, interactions are not instantaneous but are mediated by
virtual photons at the speed of light. Therefore, the Coulomb potential arises
only as leading order contribution to the electromagnetic particle-particle
interaction in a perturbation theory expansion in the fine-structure constant.
Largest QED effects are expected for the electron-nucleus potential
\cite{greiner:2009}, to which we will refer in the following exclusively and as
retardation effects of the electron-electron interactions. The latter are
however usually not as important as QED corrections of the electron-nucleus
potential. Larger corrections to the electron-electron interactions are
expected via the Breit interaction \cite{breit:1929}, which appears due to consideration of
current-current interactions and the inclusion of the correct gauge of the
interaction within a Lorentz invariant framework. Photon-frequency dependent
retardation corrections to the Breit interaction are usually less important
than QED corrections of the electron-nucleus potential considered in this work.
It is expected that QED corrections reduce the total relativistic
effects by about 1~\% in heavy atom containing-molecules \cite{indelicato:2011,pyykko:2012}. For
the hydrogen atom this QED correction of the electron-nucleus potential leads
to the so-called Lamb shift, which splits \textit{n}$\mathrm{s}_{1/2}$ and \textit{n}$\mathrm{p}_{1/2}$ 
states in one-electron systems \cite{lamb:1947}.
In the search for new physics beyond the standard model, high-precision
spectroscopy is an essential tool to test the fundamental symmetries, 
such as space- and time-reversal parity and its violations \cite{safronova:2018, berger:2019}.
Because of the numerically small effects, it is important to investigate
systems with a number of traits, which enhance effects of these violations.
Certain molecules
are particularly suited for such purposes.
One promising candidate is radium monofluoride (RaF).
It was first proposed because of its predicted nuclear-spin-dependent parity violation 
and expected suitability for laser cooling \cite{isaev:2010, Isaev:2013}.
The pear-shaped octupole deformation \cite{gaffney:2013, butler:2020} of the heavy \textsuperscript{222,224,226}Ra
isotopes enhances symmetry violation effects further.
Synthesis of the open-shell short-lived isotopically pure radioactive RaF molecules
and successive spectroscopy was first achieved with the collinear resonance ionisation spectroscopy (CRIS)
method at the ISOLDE ion-beam facility at CERN \cite{garciaruiz:2020}.
Applicable for other molecules, such as RaOH, RaO, RaH, AcF and ThO, and nuclei 
with half-lives of just milliseconds, it paved the way for high-precision spectroscopy
of radioactive isotope containing molecules.
In recent experiments, the rovibronic structure of \textsuperscript{226}Ra\textsuperscript{19}F was
obtained with a resolution, which is two orders of magnitude larger compared 
to previous experiments \cite{udrescu:2023}.
Also, eleven electronic states and excitations were reported \cite{mak:2023}.
Thus, theoretical predictions and proposals of suited molecules for precision tests
play an essential role.
However, as the experiments have reached unprecedented accuracies, and the
investigated symmetry violations are even smaller, it is necessary to develop
sophisticated theoretical frameworks further as well.
While QED corrections have been studied in atoms for a long time \cite{mohr:1974a, blundell:1993, schneider:1994, 
mohr:1998, sunnergren:1998, labzowsky:1999, artemyev:2005, karshenboim:2005, indelicato:2007, thierfelder:2010, 
shabaev:2013, schwerdtfeger:2015, ginges:2016, berengut:2016, smits:2023},
they have gained attention in molecular frameworks only recently 
\cite{koziol:2018b, sunaga:2021, skripnikov:2021, sunaga:2022, zaitsevskii:2022, colombojofre:2022, flynn:2024, flynn:2024b}.
In high-precision calculations of atoms, the leading order QED corrections
have shown to be the missing contributions to the first ionisation energy and
electron affinity of the gold atom in order to reach meV accuracy, compared
to the experimental values \cite{pasteka:2017}.
Pa\ifmmode \check{s}\else \v{s}\fi{}teka \textit{et al.} also noted, that higher excitations in the coupled cluster (CC)
method have less impact on the result, compared to the QED corrections.
Because QED contributions grow with increasing nuclear charge, as they are relativistic
effects, they are expected to contribute mainly to molecules containing heavy nuclei
such as RaF.
Overall, the effects in molecules and the influence on properties are largely unknown
and it is thus desirable to have efficient, yet accurate, options to approximate
the corrections reliably.
In this work, the leading order QED corrections are included as effective
potentials within the two-component ZORA framework.
Within this framework, one can obtain accurate results for the QED corrections,
although the overall error of ZORA, compared to four-component approaches, is in
the same order of magnitude. This incremental approach allows for an efficient 
treatment of QED effects in atoms and molecules, suited to predict size, trends and
relevance. Results of variational and perturbative treatments are compared to 
four-component variational and numerical Dirac-Hartree-Fock (DHF) calculations.
\section{Theory}

\subsection{Vacuum polarisation}
\begin{figure}[H]
\centering
\subfloat[\centering]{{\includegraphics[width=0.20\textwidth]{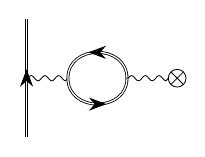}}}
\qquad
\subfloat[\centering]{{\includegraphics[width=0.20\textwidth]{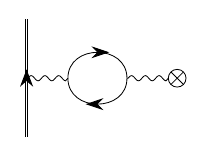}}}
\caption{Feynman diagrams of the exact bound-state VP (a) and 
Uehling potential (b). 
}
\label{figurevp}
\end{figure}

Considering instantaneous electron-positron pair creation and annihilation 
in vacuum, one obtains a temporary polarisation of the vacuum, which averages out
when one considers integration over time.
Here, the vacuum polarisation is treated as a perturbation to the 
electron-nucleus attraction term.
The schematic process is shown in Figure \ref{figurevp}a, while
\ref{figurevp}b shows the Uehling process, which assumes the
created pair to be free particles.

The Uehling potential \cite{uehling:1935} 
for the interaction between an electron and point-like nucleus with charge number $Z$ is
\begin{equation}
\label{uehlingpotential}
V_{\mathrm{u}}(\vec{r}) = 
- \frac{2\alpha}{3\pi} e \phi(\vec{r}) \int_{1}^{\infty} 
\frac{\sqrt{t^2-1}}{t^2} \left(1+\frac{1}{2t^2} \right) 
e^{-\frac{2t|\vec{r}|}{\lambdabar}} \mathrm{d}t
\end{equation}
where $\alpha$ is the fine-structure constant, 
$\phi(\vec{r})=\frac{Z}{4\pi\epsilon_0 |\vec{r}|}$ is the Coulombic
nuclear electrostatic potential with electric constant $\epsilon_0$, $\lambdabar=\frac{\hbar}{m_{\mathrm{e}}c}=\alpha a_0$ is the reduced Compton wavelength of the electron,
$r$ is the distance between electron and nucleus, $a_0$ is the Bohr radius, $\hbar$ the reduced Planck constant, $e$ the elementary charge and $m_{\mathrm{e}}$ the electron mass. 
The integral can be reformulated in terms of
modified Bessel functions \cite{frolov:2012} to obtain an
analytical expression.
Herein, however, the potential is evaluated using the method by 
Fullerton and Rinker \cite{fullerton:1976}.
Higher order corrections for the vacuum polarisation, such as 
the Wichmann-Kroll \cite{wichmann:1956} or the K\"{a}llen-Sabry 
\cite{kallen:1955} potentials are not included in this work,
since the Uehling potential is by far the largest contribution.

\subsection{Self energy}
\begin{figure}[H]
\centering
\subfloat[\centering]{{\includegraphics[width=0.20\textwidth]{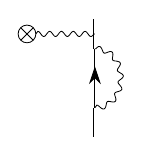}}}
\qquad
\subfloat[\centering]{{\includegraphics[width=0.20\textwidth]{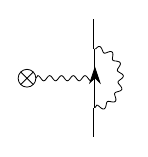}}}
\caption{Self energy (a) and vertex correction (b) 
of order $\alpha(Z\alpha)$}
\label{figurese}
\end{figure}

Other than the VP, the electron self energy is not easily written as a local potential.
The fundamental process is the exchange of a 
virtual photon of an electron with itself at a later point in time.
Figure \ref{figurese} shows the two main processes, which 
contribute to the correction. The electron is assumed to be a free particle,
for the time between emission and absorption.
Herein, the effective potentials, proposed by 
Flambaum and Ginges \cite{flambaum:2005}, as well as the 
one proposed by Pyykk\"{o} and Zhao \cite{pyykko:2003}, are considered.
The Pyykk\"{o}--Zhao potential is an exponential function with two 
fitting functions, $B_{\mathrm{pz}}(Z)$ and $\beta(Z)$, depending on the 
nuclear charge $Z$,
\begin{equation}
\label{pyykkoepotential}
V_{\mathrm{pz}}(\vec{r}) = B_{\mathrm{pz}}(Z) e^{-\beta(Z)|\vec{r}|^2}
\end{equation}
\begin{align*}
B_{\mathrm{pz}}(Z)/E_{\mathrm{h}} & = -48.6116 + 1.53666Z + 0.0301129Z^2 \\
\beta(Z)/a_{0}^{-2} & = -12751.3 + 916.038Z + 5.7797Z^2,
\end{align*}
with $E_{\mathrm{h}}$ being the Hartree energy.
The fitting functions $B_{\mathrm{pz}}(Z)$ and $\beta(Z)$ were 
obtained from 2s energy shifts of H-like systems with a finite nucleus
\cite{beier:1998, indelicato:1998} and 2s M1 hyperfine splittings of
H-like systems \cite{yerokhin:2001} and Li-like 2s states 
\cite{boucard:2000}. The H-like and Li-like reference data were
combined and yielded the potential, which can be used for \textit{n}s energy levels
and M1 hyperfine shifts of all elements $Z \geq$ 29. The lower nuclear charges were
not covered due to the reference data.
The Flambaum--Ginges potential was derived more 
rigorously \cite{flambaum:2005, berestetskii:1982} 
and splits into three contributions. Connected to the magnetic form factor
is a potential, to which we refer as the magnetic contribution $V_{\mathrm{m}}$.
From the electric form factor there arise two potentials, which we denote as the high- 
and low-frequency contributions, $V_{\mathrm{h}}$ and $V_{\mathrm{l}}$. The magnetic 
contribution for a point-like nucleus reads
\begin{equation}
\label{magneticpotential}
\begin{split}
V_{\mathrm{m}}(\vec{r}) &= -e \frac{\alpha \lambdabar}{4\pi} \mathrm{i} \vec{\bm{\gamma}}\cdot\vec{\nabla}
\left[\phi(\vec{r}) \left( \int_{1}^{\infty} \frac{e^{-\frac{2t|\vec{r}|}{\lambdabar}}}{t^2\sqrt{t^2 -1}} 
\mathrm{d}t -1 \right) \right]  \\
& = -e \frac{\alpha \lambdabar}{4\pi}\mathrm{i}\vec{\bm{\gamma}}\cdot\left[\left(\vec{\nabla}\phi(\vec{r})\right)\left(
\int_{1}^{\infty} \frac{e^{-\frac{2t|\vec{r}|}{\lambdabar}}}{t^2 \sqrt{t^2 -1}} \mathrm{d}t -1 \right)\right. \\
& \left. \quad - \phi(\vec{r}) \frac{2\vec{r}}{\lambdabar r}\int_{1}^{\infty} \frac{e^{-\frac{2t|\vec{r}|}{\lambdabar}}}{t \sqrt{t^2 -1}} \mathrm{d}t
\right] \\
& = e \phi(\vec{r}) \frac{\alpha \lambdabar}{4\pi}\mathrm{i}\vec{\bm{\gamma}}\cdot\frac{\vec{r}}{r^2} \left[
\int_{1}^{\infty} \frac{e^{-\frac{2t|\vec{r}|}{\lambdabar}}}{t^2 \sqrt{t^2 -1}} \mathrm{d}t -1 \right. \\
& \left. \quad + \frac{2r}{\lambdabar}\int_{1}^{\infty} \frac{e^{-\frac{2t|\vec{r}|}{\lambdabar}}}{t \sqrt{t^2 -1}} \mathrm{d}t
 \right] \\
\end{split}
\end{equation}
Only for a point like nucleus, $- \frac{\vec{r}}{r^2} \phi(\vec{r}) = \nabla\phi(\vec{r})$.
Compared to the Uehling potential, the magnetic potential does not just include an integral, for which no analytical solution is available,
but also the differential operator acting on the product of the integral and the
nuclear potential. However, this contribution only shows a dependence on the nuclear 
charge in terms of the nuclear model. Here, the second equation of \ref{magneticpotential} was 
implemented. In comparison, the high frequency contribution reads
\begin{equation}
\label{highfreqpotential}
\begin{split}
V_{\mathrm{h}}(\vec{r}) = & A(Z,r)\frac{\alpha}{\pi} 
e \phi(\vec{r}) \int_{1}^{\infty} 
\frac{ e^{-\frac{2t|\vec{r}|}{\lambdabar}} }{\sqrt{t^2-1}} 
\left[ \left(1-\frac{1}{2t^2}\right) \right. \\
& \times\left. \left( \ln(t^2-1) + 4\ln\left( \frac{1}{Z\alpha} + 
\frac{1}{2} \right) \right) + \frac{1}{t^2} - \frac{3}{2} 
\right] \mathrm{d}t \\
\end{split}
\end{equation}
In equation (\ref{highfreqpotential}), Flambaum and Ginges included
the  cut--off function $A(Z,r)$, with
\begin{equation}
\label{flambaumfactor}
\begin{aligned}
A(Z,r) = & \Theta(Z,\vec{r})(1.071-1.976a^2 \\
& -2.128a^3 + 0.169a^4) \\
\Theta(Z,r) = & \frac{|\vec{r}|}{|\vec{r}|+0.07(Z\alpha)^2 \lambdabar} \\
a = & (Z-80)\alpha, \\
\end{aligned}
\end{equation}
which was obtained from fitting radiative shifts for high Coulomb s-levels,
in order to account for the potential near the core. Starting from a free-particle
formulation, the potential would otherwise not be suited for small distances in
heavy atoms $Z>80$ \cite{flambaum:2005, mohr:1992a, mohr:1992b}.
Thierfelder and Schwerdtfeger \cite{thierfelder:2010} 
introduced the expression
\begin{equation}
A_{n}(Z)=A_{n0} + A_{n1} \frac{Z}{1+ \mathrm{exp}
\left[ \left( Z/A_{n2} \right)^5 \right]}
\label{schwerdtfegerfactor}
\end{equation}
depending on the main quantum number $n$ and the values of 
$A_{ni}$ were obtained from fitting self energy contributions to 
H-like atoms, calculated by Mohr \cite{mohr:1974a, mohr:1974b, mohr:1992b, 
mohr:1992a}. Multiplied with the cut--off term $\Theta(Z,r)$, this gives 
another prefactor, which is important for comparing results. 
Finally, the low frequency contribution
\begin{equation}
\label{lowfreqpotential}
V_{\mathrm{l}}(\vec{r})/E_{\mathrm{h}} = B(Z) Z^4 \alpha^3 
e^{-\frac{Z|\vec{r}|}{a_0}}
\end{equation}
includes a fitting function 
\begin{equation}
B(Z)=0.074+0.35Z\alpha
\end{equation}
which reproduces radiative shifts for the high Coulomb p-levels \cite{mohr:1992b, mohr:1992a}.
Flambaum and Ginges chose it to be in the range of the size of the 1s orbital $a_0/Z$ 
\cite{flambaum:2005}.

\subsection{Perturbative computation of QED corrections}

As QED corrections are always small compared to the total energy, it is generally sufficient to include them perturbatively. For a variational wave function the QED correction can be simply obtained as expectation value of the QED potential

\begin{equation}
\Delta E_{\mathrm{QED}} = \Braket{\Psi|V_{\mathrm{VP}}+V_{\mathrm{SE}}|\Psi},
\end{equation}

where closed expression for the VP and SE corrections to the potential energy $V_{\mathrm{VP}}$, $V_{\mathrm{SE}}$ are given above. Note that non-local SE corrections as e.g. suggested in Ref.~\cite{shabaev:2013}, would require a perturbative correction of the wave function as well.

In a mean-field approach based on the expansion of the Hilbert space in a set
of molecular orbitals $\psi_{i}$, which are represented as linear combination of
Gaussian basis functions $\chi_{\mu}$, $\psi_{i}=\sum_{\mu} C_{\mu i}\chi_{\mu}$ with
the molecular orbital or Kohn-Sham orbital coefficients $C_{\mu i}$, the QED
correction to the total energy is then obtained as the sum over all occupied
orbital contributions $i$

\begin{equation}
\Delta E_{\mathrm{QED}} = \sum\limits_i \Braket{\psi_i|V_{\mathrm{VP}}+V_{\mathrm{SE}}|\psi_i}
\end{equation}

In such a mean-field approach, first order corrections to orbital energies
$\epsilon_i$ require the evaluation of the linearly perturbed Fock matrix i.e.

\begin{equation}
\label{eq:linearresponse}
\Delta \epsilon_{i,\mathrm{QED}} = \Braket{\psi_i|V_{\mathrm{VP}}+V_{\mathrm{SE}}|\psi_i} + G_{ii}(\bm{D}'_{\mathrm{QED}})\,.
\end{equation}

The first order perturbed two-electron matrix $\bm{G}$ is defined as

\begin{multline}
G_{ii} = \sum\limits_{\mu\nu\rho\sigma}C_{\mu i}^\dagger C_{\nu i}
\left[
  D'_{\mathrm{QED},\rho\sigma}\left( (\mu\nu|\rho\sigma)
   -a_{\mathrm{X}}\frac{1}{2}(\mu\sigma|\rho\nu) \right)
\right.\\\left.
    +a_{\mathrm{DFT}}\Braket{\chi_\mu|V'_{\mathrm{XC}}(\bm{D},\bm{D}'_{\mathrm{QED}})|\chi_\nu}
\right]
\end{multline}
where the Mulliken notation for two electron integrals is employed:
$(\mu\nu|\rho\sigma)=\iint\mathrm{d}^{3}r_{1}\mathrm{d}^{3}r_{2}\chi_{\mu}(\vec{r}_{1})\chi_{\rho}(\vec{r}_{2})\frac{1}{\left|\vec{r}_{1}-\vec{r}_{2}\right|}\chi_{\nu}(\vec{r}_{1})\chi_{\sigma}(\vec{r}_2)$,
$\bm{D}$ is the density matrix with elements $D_{\mu\nu}=\sum_{i} n_{i} C_{\mu
i}^\dagger C_{\nu i}$, with the occupation vector $\vec{n}$ and the linearly
perturbed density matrix $\bm{D}'_{\mathrm{QED}}$ being obtained from the linear
response of $\psi_{i}$ or $\bm{D}$ to the QED potentials as described in
Ref.~\cite{bruck:2023}.  In case of pure DFT (non-hybrid) we have
$a_{\mathrm{X}}=0$ and in case of pure HF we have $a_{\mathrm{X}}=1$ and
$a_{\mathrm{DFT}}=0$. $V'_{\mathrm{XC}}$ is the perturbed exchange-correlation
potential.

\subsection{Picture-change transformation within ZORA}
All QED potentials discussed in this work are one-electron
operators. Therefore, all four-component representations of
these potentials given in the previous sections can be
transformed to two-component form within zeroth order regular
approximation (ZORA) using the approach and implementation
detailed in Ref.~\cite{gaul:2020}. Here we implicitly generate the small component within ZORA from the ZORA wave function as:
\begin{equation}
\psi^{\mathrm{S}}\approx \psi^{\mathrm{S}}_{\mathrm{ZORA}} =\frac{c}{2m_{\mathrm{e}}c^2-\tilde{V}}\vec{\bm{\sigma}}\cdot\vec{\hat{p}} \psi^{\mathrm{ZORA}}
\end{equation}
For this purpose we employ the model potential $\tilde{V}$ formulation of ZORA by van
W\"ullen \cite{wullen:1998}. For details on how resulting matrix elements are
derived and computed see Ref.~\cite{gaul:2020}.  If not stated explicitly
otherwise we include picture-change corrections due to a different
normalisation of the wave function in a four component framework by applying a
renormalisation of ZORA spinors as follows:
\begin{equation}
\psi^{\mathrm{rZORA}}_{i} = \frac{\psi_{i}^{\mathrm{ZORA}}}{\sqrt{1+\Braket{\psi^{\mathrm{S}}_{i,\mathrm{ZORA}}|\psi^{\mathrm{S}}_{i,\mathrm{ZORA}}}}}
\end{equation}

Response calculations and
corrections of orbital energies were computed as detailed in
Refs.~\cite{bruck:2023,colombojofre:2022}.  

\section{Methods}

\subsection{Cubic spline interpolation}
The Uehling potential was implemented for a point-like nucleus following the approach by 
Fullerton and Rinker \cite{fullerton:1976}.
The Pyykkö--Zhao potential as well as the low frequency contribution 
to the Flambaum--Ginges potential contain analytically solvable 
integrals and, therefore, could be directly implemented as operators on 
a grid (following Ref.~\cite{gaul:2020})
according to equations (\ref{pyykkoepotential}) and (\ref{lowfreqpotential}).
To our knowledge no closed-form for the integrals appearing in 
the magnetic (\ref{magneticpotential}) 
and high frequency (\ref{highfreqpotential}) contributions to the Flambaum--Ginges 
potential exist.
Therefore,
we approximated these integrals with the help of cubic 
spline interpolation. For this purpose, the integral was 
evaluated numerically on a grid using Mathematica 11.01 \cite{mathematica:11}. 
The two integrals have finite values for $r\rightarrow 0$ and approach zero for 
$r\rightarrow \infty$. They are approximated with cubic spline interpolation based 
on 1000 logarithmically spaced grid points in the range of $10^{-10} < r/a_0 < 0.1$ each.
The first grid point was chosen to be this small, because the effective potentials
probe the electronic structure at the vicinity of the nucleus.
The mean relative errors of the spline compared to the numerically obtained values at
10000 equidistant points in the same range is $1.00\times10^{-5}$.
A similar approach was taken for the integral in the high frequency contribution. In
order to exclude the nuclear charge $Z$ from the cubic spline interpolation, the integral was written as
\begin{equation}
\begin{split}
&\int_{1}^{\infty} 
\frac{ e^{-\frac{2t|\vec{r}|}{\lambdabar}} }{\sqrt{t^2-1}} 
\left[ \left(1-\frac{1}{2t^2}\right) 
\right. \\& \left. \times 
\left( \ln(t^2-1) + 4\ln\left( \frac{1}{Z\alpha} + 
\frac{1}{2} \right) \right) + \frac{1}{t^2} - \frac{3}{2} 
\right] \mathrm{d}t \\
& = 4\ln\left(\frac{1}{Z\alpha}+\frac{1}{2}\right)\int_{1}^{\infty}\frac{ e^{-\frac{2t|\vec{r}|}{\lambdabar}} }{\sqrt{t^2-1}}
\left(1-\frac{1}{2t^2}\right) \mathrm{d}t \\
& + \int_{1}^{\infty}\frac{ e^{-\frac{2t|\vec{r}|}{\lambdabar}} }{\sqrt{t^2-1}}
\left[\left(1-\frac{1}{2t^2}\right)\ln\left(t^2 -1\right)+\frac{1}{t^2}-\frac{3}{2}\right] \mathrm{d}t \\
\end{split}
\end{equation}
Just as for the magnetic contribution, the two integrals trend to finite values
for $r\rightarrow 0$ and to zero for $r\rightarrow \infty$. Cubic spline interpolation
based on 1500 logarithmically spaced grid points in the range of $10^{-10} < r/a_0 < 0.1$
yielded a mean relative error of $<6.70\times10^{-5}$ compared to the numerically obtained 
values at 10000 equidistant points in the same range. Further documentation can be found 
in the supporting information. \\

\subsection{Computational details}

All calculations were performed using a modified version
\cite{berger:2005,nahrwold:09,isaev:2012,gaul:2020,bruck:2023,colombojofre:2022,zulch:2022}
of a quasi-relativistic program package \cite{wullen:2010} based on Turbomole
\cite{ahlrichs:1989}.
If not stated otherwise, the two-component wave functions were obtained from a complex Generalised 
Hartree--Fock (cGHF) calculation using the ZORA \cite{zora:1986, zora:1993,
zora:1994, zora:1996} framework, 
employing a model potential to alleviate the gauge
dependence of ZORA as suggested by van Wüllen \cite{wullen:1998}.
The model potential was applied with additional damping of the atomic
Coulomb contribution to the model potential \cite{liu:2002}.  The nuclei were
modelled using normalised spherical Gaussian nuclear density distributions
$\varrho_{A} \left( \vec{r} \right) = \frac{\zeta_{A}^{3/2}}{\pi ^{3/2}}
\mathrm{e}^{-\zeta_{A} \left| \vec{r} - \vec{r}_{A} \right| ^2}$ with $\zeta_{A} =
\frac{3}{2 r^2 _{\mathrm{nuc},A}}$ for the Coulomb part of the electron-nucleus
potential. The root-mean-square radius $r_{\text{nuc},A}$ was chosen as
suggested by Visscher and Dyall \cite{visscher:1997}.
Wave functions were converged until the energy change between two
consecutive self-consistent field cycles was below 
$10^{-12} E_{\mathrm{h}}$. 
In subsequent computations of QED contributions, a point-like nucleus was used.
The basis sets, which were used for the 
various calculations, are all specified in the related tables.
The contributions to the orbital energies of gold in Table 
\ref{goldatom} were calculated using the Becke 3-parameters 
hybrid exchange functional with the correlation functional by Lee, 
Yang and Parr (B3LYP) \cite{dirac:1929, slater:1951, vosko:1980, 
becke:1988, lee:1988, becke:1993}. For the perturbative treatment of QED corrections reported in Table \ref{goldatom},
the response calculation was converged to $10^{-9} E_{\mathrm{h}}$.
For a better comparison to the work of Thierfelder and Schwerdtfeger, the results of Tables 
\ref{groupsionisations}, \ref{scalingparameters} and \ref{transitions} were obtained 
using the factor of equation (\ref{schwerdtfegerfactor}) in the 
high-frequency contribution to the Flambaum--Ginges effective potential. 
For all other calculations, the factor by Flambaum and Ginges 
(\ref{flambaumfactor}) was used. If not stated otherwise, the following isotopes were used:
$^{4}_{2}$He, $^{7}_{3}$Li, $^{9}_{4}$Be, $^{11}_{5}$B, $^{20}_{10}$Ne, $^{23}_{11}$Na, $^{24}_{12}$Mg, $^{27}_{13}$Al,
$^{40}_{18}$Ar, $^{39}_{19}$K, $^{40}_{20}$Ca, $^{64}_{29}$Cu, $^{65}_{30}$Zn, $^{70}_{31}$Ga, $^{84}_{36}$Kr,
$^{85}_{37}$Rb, $^{88}_{39}$Sr, $^{91}_{40}$Zr, $^{108}_{47}$Ag, $^{112}_{48}$Cd, $^{115}_{49}$In, $^{119}_{50}$Sn,
$^{131}_{54}$Xe, $^{133}_{55}$Cs, $^{137}_{56}$Ba, $^{144}_{60}$Nd, $^{173}_{70}$Yb, $^{197}_{79}$Au, $^{201}_{80}$Hg,
$^{204}_{81}$Tl, $^{222}_{86}$Rn, $^{223}_{87}$Fr, $^{226}_{88}$Ra, $^{232}_{90}$Th.
Relative deviations (in percent) as denoted in the various tables as Dev. were calculated according
to $100\times|(X - X_{\mathrm{lit}})/X_{\mathrm{lit}}|$, where $X$ corresponds to our computed value
and $X_{\mathrm{lit}}$ to the corresponding literature value we compare to.

\section{Results and discussion}

\begin{table}
\begin{threeparttable}
\captionsetup{format=plain,singlelinecheck=false,justification=Justified,font=scriptsize}
\renewcommand{\arraystretch}{1.3}
\caption{\label{groupsionisations} SE (Flambaum--Ginges) and VP (Uehling) contributions in eV to 
the ionisation energies of group 1, 2, 11, 12, 13 and 18 atoms, calculated as expectation values 
based on ZORA-HF/dyall.ae4z calculations, using prefactor (\ref{schwerdtfegerfactor}). 
Comparison \cite{thierfelder:2010} with four-component DHF calculations with perturbative 
treatment of the QED contributions.}
\begin{tabular*}{\columnwidth}{
l
S[table-column-width=0.28\columnwidth,table-format=1.3e-2,round-precision=3,round-mode=places]
S[table-column-width=0.28\columnwidth,table-format=-1.3e-2,round-precision=3,round-mode=places]
S[table-column-width=0.15\columnwidth,table-format=1.1,round-precision=1,round-mode=places]
S[table-column-width=0.15\columnwidth,table-format=1.1,round-precision=1,round-mode=places]
}
\hline\hline
& {$V_{\mathrm{SE}}$} & {$V_{\mathrm{VP}}$} & {Dev./\%} & {Dev./\%} \\
\hline 
Li &  -3.372e-5   &  1.178e-6  &  0.1  &  0.9  \\
Na &  -3.051e-4   &  1.628e-5  &  0.3  &  0.5  \\
K  &  -5.486e-4   &  3.724e-5  &  0.1  &  0.2  \\
Rb &  -1.452e-3   &  1.443e-4  &  0.7  &  0.3  \\
Cs &  -2.389e-3   &  3.299e-4  &  0.5  &  0.5  \\
Fr &  -6.392e-3   &  1.614e-3  &  2.1  &  1.8  \\
\hline
Be\tnote{a} &  -9.019e-5   &  3.420e-6  &  0.2  &  0.9  \\
Be\tnote{b} &  -8.044e-5   &  3.046e-6  &  11.2 &  12.4 \\
Mg &  -4.424e-4   &  2.438e-5  &  0.3  &  1.3  \\
Ca &  -6.873e-4   &  4.775e-5  &  1.5  &  1.7  \\
Sr &  -1.678e-3   &  1.692e-4  &  1.6  &  2.1  \\
Ba &  -2.666e-3   &  3.724e-4  &  2.3  &  2.4  \\
Ra &  -6.842e-3   &  1.743e-3  &  2.0  &  2.6  \\
\hline
Cu &  -3.363e-3   &  2.769e-4  &  1.1  &  0.1  \\
Ag &  -7.420e-3   &  8.505e-4  &  0.6  &  0.2\tnote{b}  \\
Au &  -2.661e-2   &  5.294e-3  &  0.7  &  0.2  \\
\hline
Zn &  -3.342e-3   &  2.818e-4  &  0.7  &  1.6  \\
Cd &  -7.209e-3   &  8.434e-4  &  1.1  &  1.5  \\
Hg &  -2.566e-2   &  5.214e-3  &  0.3  &  0.9  \\
\hline
B  &   2.365e-4   &  -9.815e-6  &  0.7  &  1.2  \\
Al &   5.741e-4   &  -3.323e-5  &  0.6  &  3.5  \\
Ga &   2.130e-3   &  -1.798e-4  &  5.6  &  5.3  \\
In &   3.657e-3   &  -4.230e-4  &  5.5  &  5.5  \\
Tl &   6.530e-3   &  -1.282e-3  &  4.9  &  4.6  \\
\hline
He &  -1.776e-4   &   5.802e-6  &  0.3  &  1.2  \\
Ne &   1.032e-3   &  -5.982e-5  &  4.0  &  5.2  \\
Ar &   1.209e-3   &  -9.296e-5  &  4.0  &  5.2  \\
Kr &   2.285e-3   &  -2.908e-4  &  2.9  &  4.4  \\
Xe &   2.997e-3   &  -5.899e-4  &  2.8  &  4.0  \\
Rn &   5.622e-3   &  -2.260e-3  &  0.7  &  3.3  \\
\hline\hline
\end{tabular*}
\begin{tablenotes}
\item[a] Restricted Hartree--Fock (RHF) solution
\item[b] Lower energy broken-symmetry solution with deviations
higher than those of the RHF solution, which agrees excellently
with the literature.
\item[c] Deviation obtained from a corrected
four-component VP contribution of 
\SI{8.492e-4}{\eV}, instead of \SI{8.930e-4}{\eV} 
as obtained previously in 
Ref.~\onlinecite{thierfelder:2010}. The latter
resulted from a minor deficiency in a GRASP
routine that assumed a VP large-distance limit to be
reached once, for some distance, deviations between 
asymptotic expansion and explicit computation fall
below a given threshold. Here, however, curves
crossed rather than merged so that
a grid point coincidentally hitting this crossing flagged
erroneously an arrival in the large-distance regime.
\end{tablenotes}
\end{threeparttable}
\end{table}

\begin{figure}
\captionsetup{justification=Justified}
\includegraphics[width=\columnwidth]{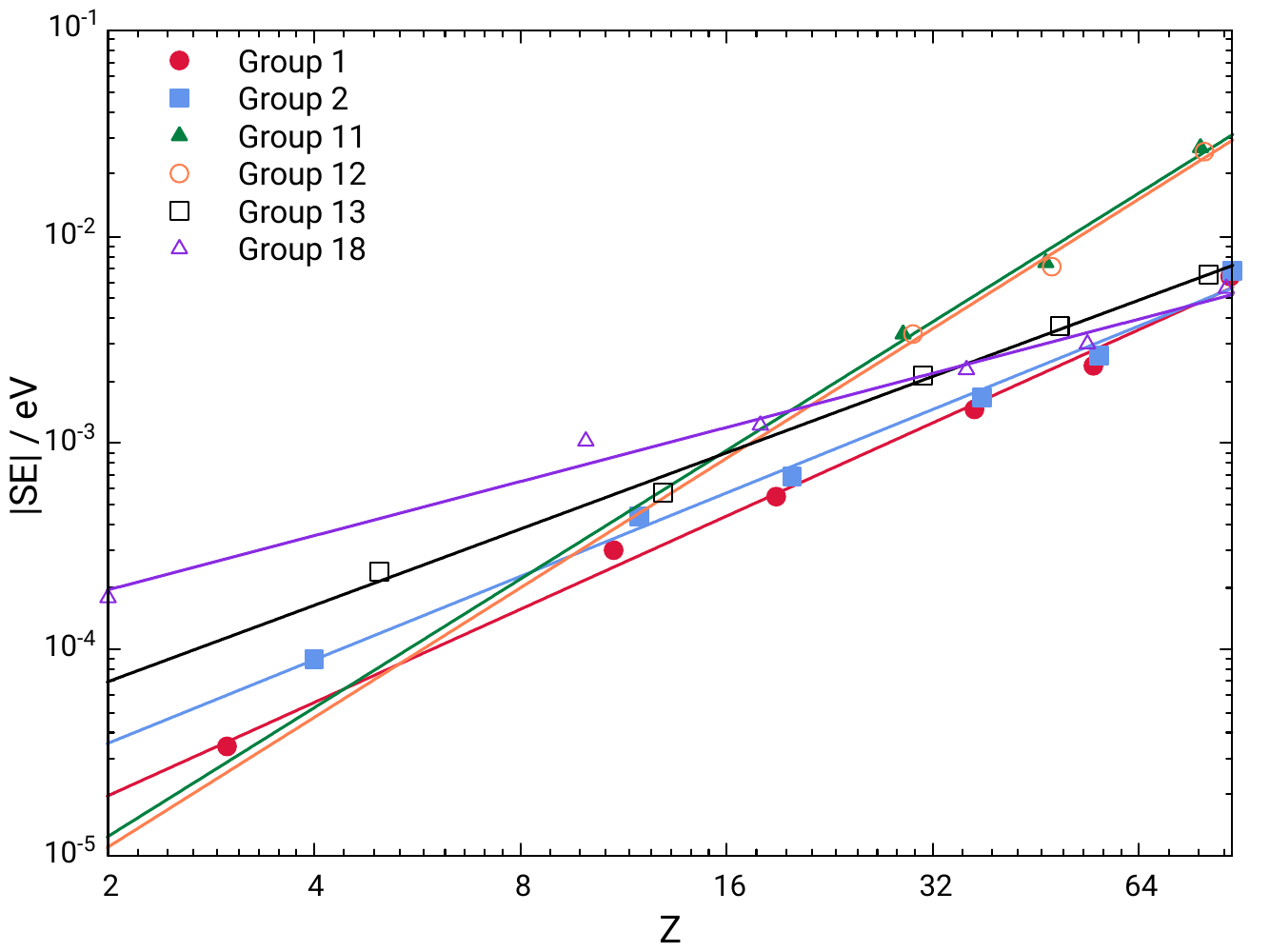}
\includegraphics[width=\columnwidth]{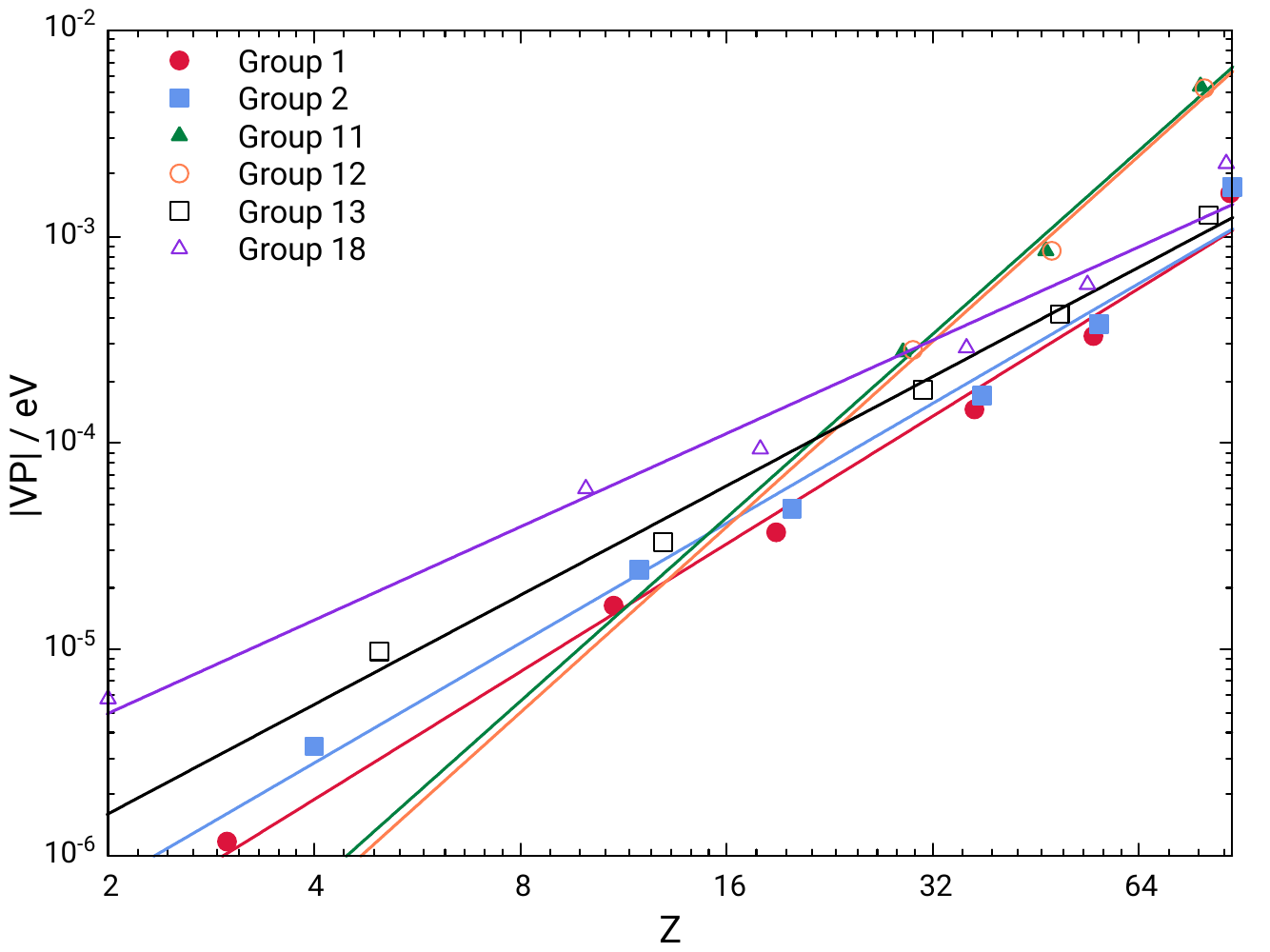}
\caption{\label{loglogplots} Absolute SE (Flambaum--Ginges) and VP (Uehling) contributions to 
the ionisation energies of group 1, 2, 11, 12 and 18 atoms from Table \ref{groupsionisations} with linear fits in a log-log scale plot.}
\end{figure}

\begin{table}
\captionsetup{format=plain,singlelinecheck=false,justification=Justified,font=scriptsize}
\renewcommand{\arraystretch}{1.3}
\caption{\label{scalingparameters} Scaling parameters for the SE (Flambaum--Ginges) and VP (Uehling) 
contributions to ionisation energies of atoms, sorted by groups according to $AZ^B$, obtained 
from linear regression of the results in Table \ref{groupsionisations}, shown in Figure \ref{loglogplots}.}
\begin{tabular*}{\columnwidth}{
c
S[table-column-width=0.26\columnwidth,table-format=-1.3e-2,round-precision=3,round-mode=places]
S[table-column-width=0.14\columnwidth,table-format=-1.3e-2,round-precision=3,round-mode=places]
S[table-column-width=0.26\columnwidth,table-format=-1.3e-2,round-precision=3,round-mode=places]
S[table-column-width=0.14\columnwidth,table-format=-1.3e-2,round-precision=3,round-mode=places]
}
\hline\hline
{Group} & {A$_{\mathrm{SE}}$} & {B$_{\mathrm{SE}}$} & {A$_{\mathrm{VP}}$} & {B$_{\mathrm{VP}}$} \\
\hline
1   & 6.91306e-6  &  1.49891 & 1.084276e-7  &  2.05442e-0  \\
2   & 1.38315e-5  &  1.34211 & 1.941780e-7  &  1.92756e-0  \\
11  & 2.95311e-6  &  2.06898 & 1.205091e-8  &  2.95158e-0  \\
12  & 2.59543e-6  &  2.08409 & 9.942144e-9  &  2.98347e-0  \\
13  & 2.97715e-5  &  1.22705 & 4.714982e-7  &  1.75795e-0  \\
18  & 1.06299e-4  &  0.87047 & 1.727229e-6  &  1.50050e-0  \\
\hline\hline
\end{tabular*}
\end{table}

\begin{table*}
\begin{threeparttable}
\captionsetup{format=plain,singlelinecheck=false,justification=Justified,font=scriptsize}
\renewcommand{\arraystretch}{1.3}
\caption{\label{transitions} SE (Flambaum--Ginges) and VP (Uehling) contributions in eV to 
the ${}^{2}\mathrm{P}_{1/2} \leftarrow {}^{2}\mathrm{S}_{1/2}$ and ${}^{2}\mathrm{P}_{3/2} \leftarrow {}^{2}\mathrm{S}_{1/2}$ transitions 
calculated as expectation values based on ZORA-HF/dyall.cv3z calculations, using prefactor (\ref{schwerdtfegerfactor}). 
Comparison \cite{thierfelder:2010} with four-component numerical 
DHF calculations with perturbative treatment of the QED contributions. $Z$ is the nuclear charge and
 $N$ is the number of electrons.}
\begin{tabular*}{2.0\columnwidth}{cc
S[table-column-width=0.28\columnwidth,table-format=-1.3e-2,round-precision=3,round-mode=places]
S[table-column-width=0.28\columnwidth,table-format=-1.3e-2,round-precision=3,round-mode=places]
S[table-column-width=0.15\columnwidth,table-format=1.1,round-precision=1,round-mode=places]
S[table-column-width=0.15\columnwidth,table-format=1.1,round-precision=1,round-mode=places]
S[table-column-width=0.28\columnwidth,table-format=-1.3e-2,round-precision=3,round-mode=places]
S[table-column-width=0.28\columnwidth,table-format=-1.3e-2,round-precision=3,round-mode=places]
S[table-column-width=0.15\columnwidth,table-format=2.1,round-precision=1,round-mode=places]
S[table-column-width=0.15\columnwidth,table-format=2.1,round-precision=1,round-mode=places]
}
\hline\hline
$Z$ & $N$ & {$V_{\mathrm{SE,2P1/2}}$} & {$V_{\mathrm{SE,2P3/2}}$}  & 
{$\mathrm{Dev./\%}$} & {$\mathrm{Dev./\%}$} & {$V_{\mathrm{VP,2P1/2}}$} & 
{$V_{\mathrm{VP,2P3/2}}$}  & {$\mathrm{Dev./\%}$} & {$\mathrm{Dev./\%}$} \\
\hline
10  &  3  &  -1.4983e-2   &  -1.4356e-2    &  0.0   &  0.0   &  7.642993e-04  &   7.643267e-04   &  0.0  &  0.0  \\
20  &  3  &  -2.0654e-1   &  -1.9358e-1    &  0.5   &  0.0   &  1.414364e-02  &   1.417365e-02   &  1.4\tnote{a}  &  0.7\tnote{b}  \\
30  &  3  &  -8.8641e-1   &  -8.1793e-1    &  0.1   &  0.1   &  7.561986e-02  &   7.616341e-02   &  0.7  &  0.7  \\
40  &  3  &  -2.4476e-0   &  -2.2375e-0    &  0.3   &  0.4   &  2.525861e-1   &   2.568468e-1    &  0.4  &  0.4  \\
50  &  3  &  -5.3793e-0   &  -4.9055e-0    &  0.5   &  0.6   &  6.631154e-01  &   6.837874e-01   &  0.3  &  0.2  \\
60  &  3  &  -1.0316e+1   &  -9.4666e-0    &  1.1   &  1.0   &  1.513493e+00  &   1.591082e+00   &  0.1  &  0.3  \\
70  &  3  &  -1.8099e+1   &  -1.6909e+1    &  1.1   &  1.3   &  3.168571e+00  &   3.420085e+00   &  0.4  &  0.6  \\
80  &  3  &  -2.9884e+1   &  -2.8847e+1    &  1.2   &  1.6   &  6.293351e+00  &   7.040060e+00   &  0.5  &  0.9  \\
90  &  3  &  -4.7252e+1   &  -4.8020e+1    &  0.9   &  1.5   &  1.212508e+01  &   1.424151e+01   &  0.0  &  0.6  \\
\hline                                                                                        
20  &  11 &  -2.9162e-2   &  -2.7515e-2    &  0.0   &  0.0   &  1.964694e-03  &   1.967590e-03   &  0.0  &  0.0  \\
30  &  11 &  -1.7364e-1   &  -1.6109e-1    &  0.0   &  0.0   &  1.449537e-02  &   1.459192e-02   &  0.7  &  0.7  \\
40  &  11 &  -5.4570e-1   &  -5.0144e-1    &  0.0   &  0.2   &  5.483776e-02  &   5.577522e-02   &  0.9  &  0.9  \\
50  &  11 &  -1.2846e-0   &  -1.1769e-0    &  0.1   &  0.0   &  1.534427e-01  &   1.584869e-01   &  1.3  &  1.3  \\
80  &  11 &  -7.7636e-0   &  -7.4541e-0    &  0.5   &  0.3   &  1.555312e+00  &   1.759592e+00   &  1.5  &  1.1  \\
90  &  11 &  -1.2391e+1   &  -1.2414e+1    &  1.1   &  0.8   &  2.999275e+00  &   3.583824e+00   &  2.4  &  1.7  \\
\hline                                                                                        
40  &  29 &  -7.9050e-2   &  -7.3224e-2    &  0.1   &  0.0   &  7.892721e-03  &   8.005152e-03   &  1.3  &  1.2  \\   
50  &  29 &  -2.6579e-1   &  -2.4449e-1    &  0.0   &  0.2   &  3.146359e-02  &   3.240273e-02   &  0.9  &  0.9  \\
60  &  29 &  -6.3158e-1   &  -5.8264e-1    &  0.2   &  0.0   &  8.837477e-02  &   9.299160e-02   &  1.0  &  0.9  \\
70  &  29 &  -1.2594e-0   &  -1.1770e-0    &  0.3   &  0.1   &  2.087889e-01  &   2.263268e-01   &  1.0  &  0.9  \\
90  &  29 &  -3.7619e-0   &  -3.7487e-0    &  1.1   &  0.6   &  8.990005e-01  &   1.072575e+00   &  2.1  &  1.5  \\
\hline\hline
\end{tabular*}
\begin{tablenotes}
\item[a] Deviation from a four-component value of \SI{1.42559e-2}{\eV}, which corrects a misprint in Ref.~\onlinecite{thierfelder:2010}.
\item[b] Deviation from a four-component value of \SI{1.42905e-2}{\eV}, which corrects a misprint in Ref.~\onlinecite{thierfelder:2010}.
\end{tablenotes}
\end{threeparttable}
\end{table*}

\begin{table}
\captionsetup{format=plain,singlelinecheck=false,justification=Justified,font=scriptsize}
\renewcommand{\arraystretch}{1.3}
\caption{\label{group1and11homo} SE (Flambaum--Ginges) and VP (Uehling) contributions in eV to 
the valence orbital energies of group 1 and 11 atoms calculated as expectation values based on 
ZORA-HF/dyall.v3z, using prefactor (\ref{flambaumfactor}). Comparison \cite{sunaga:2022} with 
four-component AOC-HF/dyall.v3z calculations based on DC Hamiltonians.}
\begin{tabular*}{\columnwidth}{
l
S[table-column-width=0.28\columnwidth,table-format=1.3e-2,round-precision=3,round-mode=places]
S[table-column-width=0.28\columnwidth,table-format=-1.3e-2,round-precision=3,round-mode=places]
S[table-column-width=0.15\columnwidth,table-format=1.1,round-precision=1,round-mode=places]
S[table-column-width=0.15\columnwidth,table-format=1.1,round-precision=1,round-mode=places]
}
\hline\hline
& {$V_{\mathrm{SE}}$} & {$V_{\mathrm{VP}}$} & {Dev./\%} & {Dev./\%} \\
\hline 
Li &    4.0244e-5   &   -1.3504e-6    &  1.66   &   1.68   \\
Na &    2.9337e-4   &   -1.5270e-5    &  0.54   &   0.59   \\
K  &    5.1329e-4   &   -3.4064e-5    &  0.43   &   0.50   \\
Rb &    1.3594e-3   &   -1.3065e-4    &  0.15   &   0.15   \\
Cs &    2.3049e-3   &   -2.9890e-4    &  0.04   &   0.00   \\
Fr &    6.3850e-3   &   -1.4517e-3    &  0.82   &   0.97   \\
Cu &    2.8512e-3   &   -2.3629e-4    &  0.39   &   0.34   \\
Ag &    6.4641e-3   &   -7.3562e-4    &  0.25   &   0.19   \\
Au &    2.3737e-2   &   -4.6369e-3    &  0.00   &   0.04   \\
\hline\hline
\end{tabular*}
\end{table}

\begin{table*}
\begin{threeparttable}
\captionsetup{format=plain,singlelinecheck=false,justification=Justified,font=scriptsize}
\renewcommand{\arraystretch}{1.3}
\caption{\label{goldatom} Perturbative and self-consistent QED contributions in $E_{\mathrm{h}}$ to 
the Kohn--Sham orbital energies of gold, calculated on ZORA-B3LYP/dyall.3zp level 
of theory, using prefactor (\ref{flambaumfactor}) and deviation from self-consistent 
four-component B3LYP/dyall.3zp calculations of Ref.~\cite{sunaga:2022} (Dev.). FG is the 
Flambaum--Ginges, PZ the Pyykkö--Zhao and UE the Uehling potential.}
\begin{tabular*}{2.0\columnwidth}{
l
S[table-column-width=0.34\columnwidth,table-format=-1.3e-2,round-precision=3,round-mode=places]
S[table-column-width=0.34\columnwidth,table-format=-1.3e-2,round-precision=3,round-mode=places]
S[table-column-width=0.34\columnwidth,table-format=-1.3e-2,round-precision=3,round-mode=places]
S[table-column-width=0.18\columnwidth,table-format=2.1,round-precision=1,round-mode=places]
S[table-column-width=0.34\columnwidth,table-format=-1.3e-2,round-precision=3,round-mode=places]
S[table-column-width=0.18\columnwidth,table-format=2.1,round-precision=1,round-mode=places]
}
\hline\hline
\multicolumn{1}{c}{} &
\multicolumn{2}{c|}{Perturbative} &
\multicolumn{4}{c}{Self-consistent} \\
& {$V_{\mathrm{FG+UE}}$\tnote{a}} & \multicolumn{1}{c|}{{$V_{\mathrm{PZ+UE}}$\tnote{a}}} 
& {$V_{\mathrm{FG+UE}}$\tnote{a}} & {Dev./\%} & {$V_{\mathrm{PZ+UE}}$\tnote{a}} & {Dev./\%} \\
\hline
1s\textsubscript{1/2}   &  8.284648e+00  &  8.278835e+00  &  8.26234126e+00  &  29.6  &  8.26695485e+00  &  32.4 \\
2s\textsubscript{1/2}   &  8.991031e-01  &  9.108763e-01  &  8.96653651e-01  &  6.14  &  9.09779793e-01  &  5.96 \\
2p\textsubscript{1/2}   &  6.655639e-02  &  6.668383e-02  &  6.79604042e-02  &  15.3  &  6.77671252e-02  &  13.8 \\
2p\textsubscript{3/2}   &  1.265163e-01  & -3.210891e-02  &  1.27494997e-01  &  6.78  & -3.12437621e-02  &  2.26 \\
3s\textsubscript{1/2}   &  1.912393e-01  &  1.967650e-01  &  1.90702905e-01  &  2.42  &  1.96519518e-01  &  2.13 \\
3p\textsubscript{1/2}   &  7.968179e-03  &  1.547967e-02  &  8.21469478e-03  &  12.2  &  1.56582916e-02  &  7.11 \\
3p\textsubscript{3/2}   &  2.251045e-02  & -8.719021e-03  &  2.26895571e-02  &  2.53  & -8.56409155e-03  &  0.56 \\
3d\textsubscript{3/2}   & -1.445197e-02  & -9.180955e-03  & -1.44071969e-02  &  0.77  & -9.16782105e-03  &  0.86 \\
3d\textsubscript{5/2}   & -5.968010e-03  & -8.759417e-03  & -5.93025764e-03  &  1.94  & -8.75126518e-03  &  1.91 \\
4s\textsubscript{1/2}   &  4.732438e-02  &  4.897051e-02  &  4.71720466e-02  &  1.44  &  4.88896916e-02  &  1.16 \\
4p\textsubscript{1/2}   &  1.254153e-03  &  3.506601e-03  &  1.29653232e-03  &  15.4  &  3.53131949e-03  &  5.85 \\
4p\textsubscript{3/2}   &  4.694869e-03  & -2.449237e-03  &  4.72186849e-03  &  1.53  & -2.42925519e-03  &  0.08 \\
4d\textsubscript{3/2}   & -3.434917e-03  & -2.293396e-03  & -3.44096426e-03  &  0.23  & -2.30764072e-03  &  0.04 \\
4d\textsubscript{5/2}   & -1.659251e-03  & -2.188095e-03  & -1.66667094e-03  &  1.15  & -2.20331053e-03  &  0.82 \\
5s\textsubscript{1/2}   &  9.148622e-03  &  9.540583e-03  &  9.10473072e-03  &  1.13  &  9.50842279e-03  &  0.83 \\
4f\textsubscript{5/2}   & -2.384824e-03  & -1.449615e-03  & -2.40348511e-03  &  0.84  & -1.47194653e-03  &  0.75 \\
4f\textsubscript{7/2}   & -1.828452e-03  & -1.416052e-03  & -1.84700332e-03  &  1.26  & -1.43838410e-03  &  0.91 \\
5p\textsubscript{1/2}   & -1.139062e-05  &  4.544576e-04  & -1.65406176e-05  &  63.8  &  4.44031672e-04  &  8.16 \\
5p\textsubscript{3/2}   &  5.803043e-04  & -5.555839e-04  &  5.72606481e-04  &  1.90  & -5.66603726e-04  &  0.65 \\
5d\textsubscript{3/2}   & -4.479991e-04  & -3.274482e-04  & -4.56800581e-04  &  2.02  & -3.39055855e-04  &  2.73 \\
5d\textsubscript{5/2}   & -2.758206e-04  & -3.040568e-04  & -2.87424975e-04  &  0.24  & -3.18480954e-04  &  0.66 \\
6s\textsubscript{1/2}   &  7.226456e-04  &  7.459826e-04  &  7.19000618e-04  &  10.3  &  7.39927313e-04  &  10.0 \\
\hline\hline
\end{tabular*}
\begin{tablenotes}
\item[a] In case two different values were obtained in the cGHF framework for a Kramers pair, contributions were averaged.
\end{tablenotes}
\end{threeparttable}
\end{table*}

\begin{table*}
\begin{threeparttable}
\captionsetup{format=plain,singlelinecheck=false,justification=Justified,font=scriptsize}
\renewcommand{\arraystretch}{1.3}
\caption{\label{aucarvgl} SE (Flambaum--Ginges) contributions in $E_{\mathrm{h}}$ to Hartree--Fock orbital energies of selected
atoms calculated as expectation values based on ZORA-HF/dyall.ae4z calculations, using prefactor (\ref{flambaumfactor}) 
and comparison to results by Kozio\l{} and Aucar \cite{koziol:2018}.}
\begin{tabular*}{2.0\columnwidth}{c
S[table-column-width=0.30\columnwidth,table-format=-1.3e-2,round-precision=3,round-mode=places]
S[table-column-width=0.30\columnwidth,table-format=-1.3e-2,round-precision=3,round-mode=places]
S[table-column-width=0.30\columnwidth,table-format=-1.3e-2,round-precision=3,round-mode=places]
S[table-column-width=0.30\columnwidth,table-format=-1.3e-2,round-precision=3,round-mode=places]
S[table-column-width=0.30\columnwidth,table-format=-1.3e-2,round-precision=3,round-mode=places]
S[table-column-width=0.30\columnwidth,table-format=-1.3e-2,round-precision=3,round-mode=places]
}
\hline\hline
\multicolumn{1}{c}{} &
\multicolumn{2}{c}{Zn} &
\multicolumn{2}{c}{Cd} &
\multicolumn{2}{c}{Hg}  \\
 &  {$V_{\mathrm{SE}}$}  &  {$\mathrm{Ref.}$\tnote{a}}  &  {$V_{\mathrm{SE}}$}  &  {$\mathrm{Ref.}$\tnote{a}} 
 &  {$V_{\mathrm{SE}}$}  &  {$\mathrm{Ref.}$\tnote{a}}  \\ 
\hline
1s\textsubscript{1/2}  &  2.83388e-1  &   2.466e-1   &  1.49908e+0  &  1.220e-0   &   1.00746e+1  &  7.408e-0  \\
2s\textsubscript{1/2}  &  2.70958e-2  &   2.613e-2   &  1.58677e-1  &  1.501e-1   &   1.21843e+0  &  1.130e-0  \\
2p\textsubscript{1/2}  & -1.38539e-4  &  -4.74e-4    &  3.91694e-3  & -2.0e-5     &   1.62551e-1  &  9.3970e-2 \\
2p\textsubscript{3/2}  &  1.77820e-3  &   1.18e-3    &  1.61787e-2  &  1.114e-2   &   1.85704e-1  &  1.296e-1  \\
3s\textsubscript{1/2}  &  3.86977e-3  &   3.815e-3   &  2.99148e-2  &  2.920e-2   &   2.68580e-1  &  2.615e-1  \\
3p\textsubscript{1/2}  & -2.38652e-5  &  -3.7e-5     &  6.09593e-4  &  3.2e-4     &   3.29344e-2  &  2.689e-2  \\ 
3p\textsubscript{3/2}  &  2.14223e-4  &   1.5e-4     &  2.78384e-3  &  2.11e-3    &   3.89315e-2  &  3.112e-2  \\
3d\textsubscript{3/2}  & -2.04187e-5  &  -7.0e-6     & -2.65978e-4  & -1.7e-4     &  -1.20076e-3  & -1.83e-3   \\
3d\textsubscript{5/2}  &  2.29478e-5  &   7.0e-6     &  4.62734e-4  &  2.1e-4     &   7.00890e-3  &  3.690e-3  \\
4s\textsubscript{1/2}  &  1.58786e-4  &   1.6e-4     &  5.35382e-3  &  5.281e-3   &   6.68059e-2  &  6.589e-2  \\
4p\textsubscript{1/2}  &              &              &  9.49469e-5  &  6.6e-5     &   7.83265e-3  &  6.854e-3  \\
4p\textsubscript{3/2}  &              &              &  4.34512e-4  &  3.3e-4     &   9.18567e-3  &  7.493e-3  \\
4d\textsubscript{3/2}  &              &              & -2.46155e-5  & -1.e-5      &  -1.19899e-4  & -2.7e-4    \\
4d\textsubscript{5/2}  &              &              &  4.66454e-5  &  2.0e-5     &   1.58518e-3  &  7.31e-4   \\
\hline\hline
\end{tabular*}
\end{threeparttable}
\begin{tablenotes}
\item[a] Additional digits, exceeding those reported in Ref.~\onlinecite{koziol:2018}, were provided 
         by K.~Kozio\l{} in personal communication.
\end{tablenotes}
\end{table*}

\begin{table}
\captionsetup{format=plain,singlelinecheck=false,justification=Justified,font=scriptsize}
\renewcommand{\arraystretch}{1.3}
\caption{\label{monofluoridesionisations} SE (Flambaum--Ginges) and VP (Uehling) contributions in eV 
to the ionisation energies of group 2 monofluorides calculated as expectation values based on 
ZORA-HF/dyall.cv3z calculations, using prefactor (\ref{flambaumfactor}). 
The bond lengths for the neutral molecule $r_0$ and the ion $r_+$ are in pm.}
\begin{tabular*}{\columnwidth}{
l
S[table-column-width=0.13\columnwidth]
S[table-column-width=0.13\columnwidth]
S[table-column-width=0.29\columnwidth,table-format=-1.3e-2,round-precision=3,round-mode=places]
S[table-column-width=0.29\columnwidth,table-format=-1.3e-2,round-precision=3,round-mode=places]
}
\hline\hline
& {$r_0$} & {$r_+$} & {$V_{\mathrm{FG}}$} & {$V_{\mathrm{UE}}$} \\
\hline
BeF  &  135.39   &   130.10  &  -1.4704528e-5   &   -2.4064935e-7   \\
MgF  &  173.63   &   167.26  &  -3.4220974e-4   &    1.8780369e-5   \\
CaF  &  198.01   &   189.49  &  -6.2846131e-4   &    4.3583886e-5   \\
SrF  &  210.26   &   201.66  &  -1.6378377e-3   &    1.6308413e-4   \\
BaF  &  220.51   &   211.95  &  -2.8343891e-3   &    3.8155570e-4   \\
RaF  &  227.73   &   219.43  &  -7.2736715e-3   &    1.7872898e-3   \\
\hline\hline
\end{tabular*}
\end{table}

\begin{table}
\captionsetup{format=plain,singlelinecheck=false,justification=Justified,font=scriptsize}
\renewcommand{\arraystretch}{1.3}
\caption{\label{bafraf} SE (Flambaum--Ginges) and VP (Uehling) contributions in $\mathrm{cm^{-1}}$ to 
the transition energies of BaF and RaF calculated as expectation values based on ZORA-HF/dyall.cv3z calculations, 
using prefactor (\ref{flambaumfactor}).}
\begin{tabular*}{\columnwidth}{
l
S[table-column-width=0.2\columnwidth,table-format=-1.3e-2,round-precision=3,round-mode=places]
S[table-column-width=0.2\columnwidth,table-format=-1.3e-2,round-precision=3,round-mode=places]
S[table-column-width=0.2\columnwidth,table-format=-1.3e-2,round-precision=3,round-mode=places]
S[table-column-width=0.2\columnwidth,table-format=-1.3e-2,round-precision=3,round-mode=places]
}
\hline\hline
\multicolumn{1}{c}{} &
\multicolumn{2}{c}{$\Pi_{1/2}\leftarrow \Sigma_{1/2}$} &
\multicolumn{2}{c}{$\Pi_{3/2}\leftarrow \Sigma_{1/2}$} \\
& {BaF} & {RaF} & {BaF} & {RaF} \\
\hline
cGHF   &   1.1559e+4    &  1.2698e+4   &   1.2110e+4   &  1.4324e+4   \\
VP     &   3.505e+0     &  1.5032e+1   &   3.562e+0    &  1.6054e+1   \\
SE     &  -2.64799e+1   & -6.17219e+1  &  -2.54324e+1  & -6.242807e+1 \\
QED    &  -2.29749e+1   & -4.6898e+1   &  -2.1870e+1   & -4.6374e+1   \\
Total  &   1.1536e+4    &  1.2651E+4   &   1.2088e+4   &  1.4277e+4   \\
\hline\hline
\end{tabular*}
\end{table}

\begin{figure}
\captionsetup{justification=Justified}
\includegraphics[width=\columnwidth]{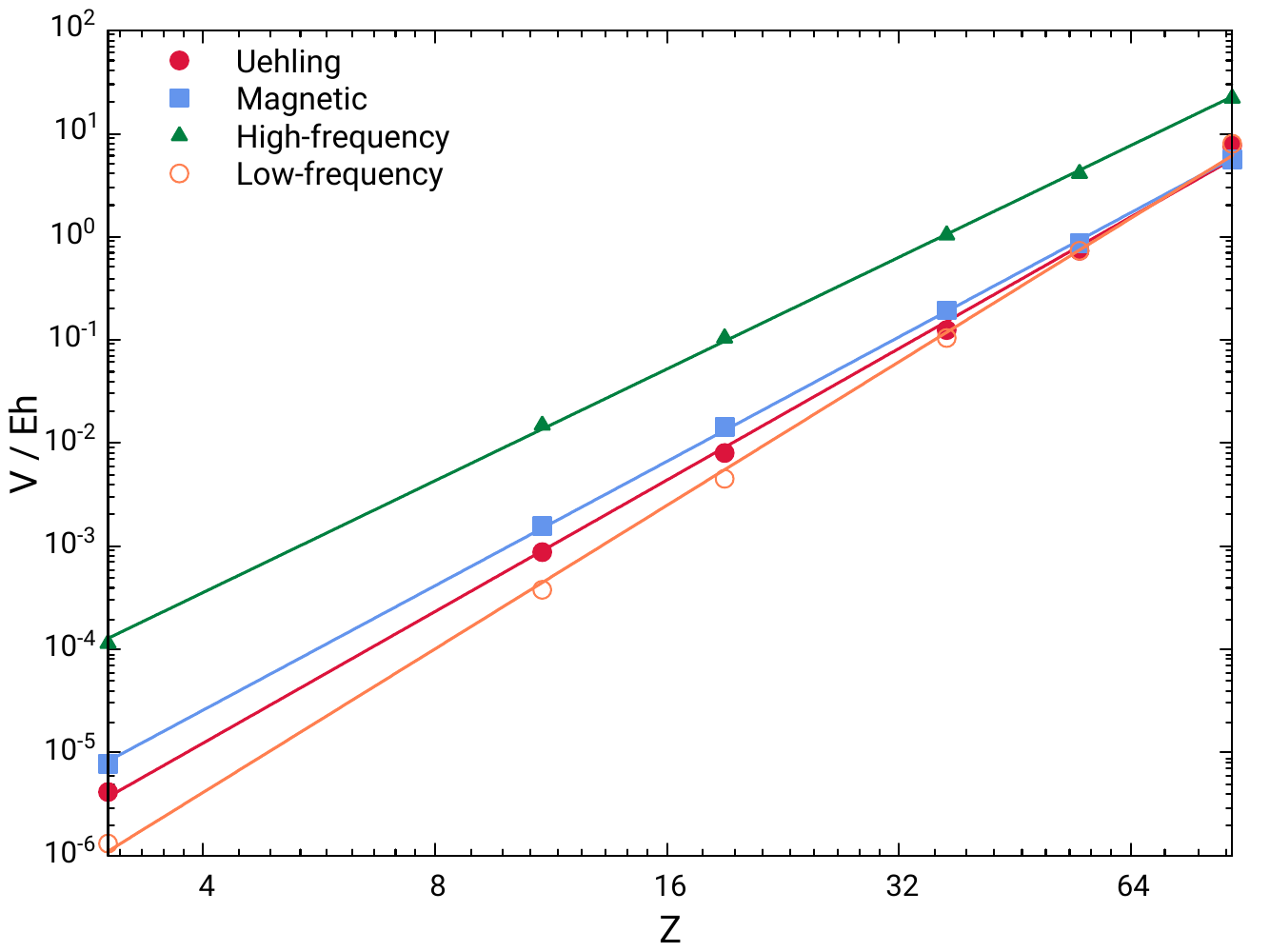}
\caption{\label{group1loglogplot} Absolute SE (Flambaum--Ginges) and VP (Uehling) contributions to 
the energies of group 1 atoms from Table (\ref{group1loglogtable}) and (\ref{group1scalingvalues}) with 
linear fits in a log-log scale plot.}
\end{figure}

\begin{table}[h]
\captionsetup{format=plain,singlelinecheck=false,justification=Justified,font=scriptsize}
\renewcommand{\arraystretch}{1.3}
\caption{\label{group1loglogtable} Scaling parameters for the SE (Flambaum--Ginges) and VP (Uehling) 
contributions to total energies of group one atoms calculated as expectation values based on ZORA-HF/dyall.cv3z 
calculations, using prefactor (\ref{flambaumfactor}). The parameters, according to $AZ^B$, were obtained 
from linear regression, shown in Figure \ref{group1loglogplot}.}
\begin{tabular*}{\columnwidth}{
c
S[table-column-width=0.22\columnwidth,table-format=-1.3e-2,round-precision=3,round-mode=places]
S[table-column-width=0.22\columnwidth,table-format=-1.3e-2,round-precision=3,round-mode=places]
S[table-column-width=0.22\columnwidth,table-format=-1.3e-2,round-precision=3,round-mode=places]
S[table-column-width=0.22\columnwidth,table-format=-1.3e-2,round-precision=3,round-mode=places]
}
\hline\hline
{} & {$V_{\mathrm{u}}$} & {$V_{\mathrm{m}}$} & {$V_{\mathrm{h}}$} & {$V_{\mathrm{l}}$} \\
\hline
A   &  3.56016e-8  &  1.02197e-7  &  2.46434e-6  &  6.94928e-9  \\
B   &  4.22608e-0  &  3.99593e-0  &  3.59070e-0  &  4.61170e-0  \\
\hline\hline
\end{tabular*}
\end{table}

\begin{table}
\captionsetup{format=plain,singlelinecheck=false,justification=Justified,font=scriptsize}
\renewcommand{\arraystretch}{1.3}
\caption{\label{group1scalingvalues} SE (Flambaum--Ginges) contributions in $E_{\mathrm{h}}$ to 
the total energies of group 1 atoms, calculated as expectation values based on 
ZORA-HF/dyall.cv3z, using prefactor (\ref{flambaumfactor}).}
\begin{tabular*}{\columnwidth}{
l
S[table-column-width=0.20\columnwidth,table-format=1.3e-2,round-precision=3,round-mode=places]
S[table-column-width=0.20\columnwidth,table-format=1.3e-2,round-precision=3,round-mode=places]
S[table-column-width=0.20\columnwidth,table-format=-1.3e-2,round-precision=3,round-mode=places]
S[table-column-width=0.20\columnwidth,table-format=-1.3e-2,round-precision=3,round-mode=places]
}
\hline\hline
 & {$V_{\mathrm{u}}$} & {$V_{\mathrm{m}}$} & {$V_{\mathrm{h}}$} & {$V_{\mathrm{l}}$} \\
\hline 
Li   & -4.13662e-6  &   7.6904e-6   &  1.1440e-4   & 1.3485e-6  \\
Na   & -8.87114e-4  &   1.5934e-3   &  1.5154e-2   & 3.8375e-4  \\
K    & -8.17309e-3  &   1.4031e-2   &  1.05896e-1  & 4.6058e-3  \\
Rb   & -1.26474e-1  &   1.8929e-1   &  1.05761e-0  & 1.0363e-1  \\
Cs   & -7.15842e-1  &   8.8586e-1   &  4.14680e-0  & 7.2203e-1  \\
Fr~~~   & -7.50459e-0  &   5.5649e-0   &  2.15144e+1  & 8.0013e-0  \\
\hline\hline
\end{tabular*}
\end{table}


\subsection{Ionisation energies}
QED contributions to the ionisation energies of group 
1, 2, 11, 12, 13 and 18 atoms are listed in Table 
\ref{groupsionisations}.
SE and VP show opposite 
sign and grow larger in magnitude with increasing nuclear charge $Z$.
For atoms up to the fifth period of the periodic table, the absolute value of the vacuum polarisation contribution is 
roughly one order of magnitude smaller than that of the electron self energy term.
In the case of higher nuclear charges, $Z > 80$, the vacuum 
polarisation contributes to the same order of magnitude.
The two contributions essentially cancel each other out at one
point, after which the VP will be greater in magnitude than the SE.
The largest absolute values of SE contributions were found for Au and Hg, followed
by their lighter group homologues Ag and Cd. The smallest
contributions, in magnitude, were computed for Li and Be. The strongest VP
contributions were also found for Au and Hg, followed by Rn and Ra.
The weakest contributions were again obtained for Li and Be. 
To determine the scaling of QED contributions to the
ionisation energies with $Z$, we plotted them with logarithmic
scales on abscissa and ordinate in Figure \ref{loglogplots}.
Linear fits for each considered group of the periodic table
of elements yielded the parameters in Table \ref{scalingparameters}.
The VP shows generally larger exponents $B$, which is in
accordance with the results obtained. QED contributions
grow faster with increasing nuclear charge for groups 11 and 12,
which points at the already known relative maximum of 
relativistic effects at group 11 \cite{pyykko:2012}.
Deviations of scaling parameters from those reported by Thierfelder
and Schwerdtfeger stem from the exclusion of super-heavy 
elements in our present work. 

Contributions to ionisation energies of group 1 and 2 atoms 
agree excellently with the four-component DHF calculations by 
Thierfelder and Schwerdtfeger \cite{thierfelder:2010} 
with deviations remaining in the range of 0.1--2.6\%.
For Be, the first result shows a deviation of 0.2\% for the SE
and 0.9\% for the VP. In addition,
we obtained another Hartree--Fock solution for Be, which is $3.2\times 10^{-4}\,E_{\mathrm{h}}$
lower in total energy. 
Using the energetically lower wave function for the neutral Be atom, 
the SE contribution weakens to $-8.044\times 10^{-5}$ eV and shows a 
deviation of 11.2\%, while the VP is lowered to
$3.046\times 10^{-6}$ eV with a deviation of 12.4\%.
The first result, which is in agreement with the literature value 
\cite{thierfelder:2010} is an RHF solution, while the second result
is a GHF solution.
Groups 11 and 12 show deviations of only 0.1--1.6\% from four-component results, 
whereas Groups 13 and 18 atoms show deviations of 0.3--5.6\%.
Except for helium, which also gives opposite signs compared
to the rest of the group 13 and 18 atoms, an 
ionisation of the p-shell instead of the s-shell was considered. 

For group 13 the electron was removed from the $\mathrm{p}_{1/2}$ 
and for group 18 from the $\mathrm{p}_{3/2}$ orbitals.
Extensive data on converged wave functions for different 
ionisations can be found in the supporting information.
It is worth noting, that the high-frequency part of
the Flambaum--Ginges potential causes the (in magnitude) largest contribution
to the SE ($\approx$ 59\% for Au, $\approx$ 71\% for Ag,
$\approx$ 79\% for Cu) and claims an even greater share of the contribution
to the ionisation energy ($\approx$ 70\% for Au, $\approx$
 78\% for Ag, $\approx$ 82\% for Cu). Consequently, the prefactor
(\ref{schwerdtfegerfactor}), introduced by Thierfelder and Schwerdtfeger,
is important to consider when comparing results.
When neglected, the computed QED corrections show greater deviations from the four-component data.
As it scales with $Z^4$, the low-frequency contribution
to the ionisation energies grows from $\approx$ 4.4\% for Cu,
over $\approx$ 6.9\% for Ag, to $\approx$ 13.2\% for Au.
The magnetic contribution shows a smaller increase ($\approx$
16.4\% for Au, $\approx$ 15.3\% for Ag, $\approx$ 13.1\% for Cu).

\subsection{Electronic transition energies}
SE and VP contributions to ${}^{2}\mathrm{P}_{1/2} \leftarrow {}^{2}\mathrm{S}_{1/2}$ 
and ${}^{2}\mathrm{P}_{3/2}\leftarrow {}^{2}\mathrm{S}_{1/2}$ transition energies of Li-, Na- 
and Cu-like ions with nuclear charges $Z=10,20,...,90$ are listed 
in Table \ref{transitions}.
As observed for the contributions to the ionisation energies, the SE and VP show 
opposite sign, with SE being roughly one order of magnitude larger in absolute value than the VP.
Comparing the same elements, Li-like ions show larger contributions 
to the transition energies than the corresponding Na- and Cu-like ions.
This aligns with the general trend of the transition energies, which 
for Li-like ions exceed those of Cu-like ions by far.
Except for the case of $\mathrm{Th^{87+}}$,
the SE contributions to the ${}^{2}\mathrm{P}_{1/2}\leftarrow {}^{2}\mathrm{S}_{1/2}$ transition energies
are larger in magnitude than the ones to the ${}^{2}\mathrm{P}_{3/2}\leftarrow {}^{2}\mathrm{S}_{1/2}$ transitions,
although the transition energy itself is much smaller in the ${}^{2}\mathrm{P}_{1/2}\leftarrow {}^{2}\mathrm{S}_{1/2}$
case. Without the exception for $\mathrm{Th^{87+}}$, the trend is the opposite for the VP.
Here, the contributions to the ${}^{2}\mathrm{P}_{3/2}\leftarrow {}^{2}\mathrm{S}_{1/2}$ transition energies are
always larger. The values obtained in the current work are compared to four-component DHF calculations
by Thierfelder and Schwerdtfeger \cite{thierfelder:2010}.
Our present SE contributions deviate from these by only 0.0--1.6\%, while the VP contributions
differ by 0.0--2.4\%.
 As for the comparison of the contributions to the 
ionisation energies in the previous section, it was crucial to use the 
prefactor of equation (\ref{schwerdtfegerfactor}). The most pronounced QED
contributions of Table \ref{transitions} are the ones to the ${}^{2}\mathrm{P}_{1/2}\leftarrow {}^{2}\mathrm{S}_{1/2}$
transition in $\mathrm{Th^{87+}}$, where they amount to $\approx$ \SI{-35}{\eV}.
They constitute a significant correction to the $\approx$ \SI{275}{\eV} transition energy.
Blundell \cite{blundell:1993} obtained a QED contribution to the ${}^{2}\mathrm{P}_{1/2}\leftarrow {}^{2}\mathrm{S}_{1/2}$
transition in $\mathrm{Th^{87+}}$ of \SI{-38.35}{\eV}, lowering the transition energy from $\approx$ \SI{309}{\eV} to
$\approx$ \SI{271}{eV}. 

\subsection{Valence orbital energies}
QED contributions to Hartree--Fock valence orbital energies of group 1 and
11 atoms are listed in Table \ref{group1and11homo}. Again, VP and SE 
contributions show opposite signs and the VP is about one order of 
magnitude smaller in absolute value than the SE. For gold, the QED corrections lift up the
 orbital energy of the HOMO by $\approx$ 0.25\%. The values compare well to
four-component average of configuration (AOC-HF) calculations reported by Sunaga,
Salman and Saue \cite{sunaga:2022}. Deviations range between 0.0\% and 1.7\%, 
although a different HF method was used. Overall, it was observed, that 
the choice of the basis set and resolution of the integration grid had 
a larger effect on the results than the level of theory.

On the density functional theory level, individual QED contributions to Kohn--Sham orbital energies of gold are
listed in Table \ref{goldatom}. The perturbative and self-consistent treatment
of the Flambaum--Ginges and Uehling (FG+UE), as well as the Pyykkö--Zhao and
Uehling potentials (PZ+UE) are compared. 
In case of two different values for the Kramers pair in the cGHF frame, the orbital 
energy contributions have been averaged.
The QED effects contribute mostly to the 
core orbitals and get smaller for the valence orbitals.
Orbitals of higher angular momentum, 
such as d- and f-orbitals, show smaller contributions compared to 
s-orbitals of the same principal quantum number. In the perturbative treatment 
of the Pyykkö--Zhao and Uehling potentials, the contributions to the d- 
and f-orbitals are significantly smaller. This is the result of the SE 
potential, which is a simple Gaussian, including two fitting functions 
for s-levels \cite{pyykko:2003}. For s-contributions, the Pyykkö--Zhao and
Flambaum--Ginges give comparable results. However, for p-, d- and f-orbitals,
the contributions differ significantly and also often show opposite signs.
The values are compared to four-component
calculations by Sunaga, Salman and Saue \cite{sunaga:2022}, who used the
same density functional and basis set. 
Our results for the self-consistent inclusion of the QED contributions
show deviations between 0.0--32.4\% compared to the reference. In both cases, the very first energy
levels show larger deviations of 2.3--32.4\%.
These are caused by the energy shift
of the ZORA Hamiltonian and mainly affect the first two energy levels.
The 6s\textsubscript{1/2} orbital contributions show deviations of 10.0\% and 10.3\%,
which could be a result of the different open-shell treatments, which in our cGHF treatment account for
spin-polarisation effects.
The \textit{n}p\textsubscript{1/2} show higher deviations compared to the remaining p-, d- and 
f-level contributions. Especially the 5p\textsubscript{1/2} of the Flambaum--Ginges
contribution shows a large deviation of 63.8\%. The origin of this deviation is presently open.

For comparison, QED corrections to
Kohn-Sham orbital energies were also calculated perturbatively. As the corrections are relatively
small, the two approaches, perturbative and self-consistent, are expected to yield comparable results. Sunaga \textit{et al.}, however,
reported noticeable differences between the two treatments on the orbital energy level.
When comparing instead the total expectation value of the (sum of) one-electron operators, virtually identical result are obtained.
If one accounts properly for the linear response as per Eq.~\ref{eq:linearresponse}, the differences between the two treatments become negligible also on the level of orbital energies.
Only the 5p\textsubscript{1/2} values stand out, because they show larger deviation between perturbative 
and self-consistent treatment of the Flambaum--Ginges term.\\
Another orbital contribution comparison is shown in Table \ref{aucarvgl}. The SE contributions, evaluated 
as orbital expectation values (without response contributions), to group 12 atoms up to the 4d shell are compared to the results reported by Kozio\l{} and Aucar \cite{koziol:2018},
who use a quite different approach to account for SE corrections (Welton picture).
Nevertheless, the s-contributions compare well and show deviations up to 14.9\% for the 1s level of Zn, 22.9\%
for Cd and 35.9\% for the 1s level of Hg. The \textit{n}p and \textit{n}d levels show larger deviations
and in the case of 2p\textsubscript{1/2} of Cd a different sign.


\subsection{Total energies}

Figure \ref{group1loglogplot} shows 
QED contributions to the total energies of group 1 atoms due to the Uehling potential and the individual terms of the Flambaum--Ginges
potential. Corresponding scaling parameters are listed in Table \ref{group1loglogtable}.
The high-frequency contributions grows fastest, whereas the low-frequency term grows slowest with
increasing the nuclear charge. The magnetic contribution scales higher than the Uehling potential. All of the
terms scale much higher for the total energy contributions compared to the corrections to the 
ionisation energies of group one atoms of Table \ref{scalingparameters}.

\subsection{Ionisation energies of group 2 monofluorides}
QED contributions to the ionisation energies of group 2 monofluorides are
listed in Table \ref{monofluoridesionisations}. The neutral molecules, as well
as the ions, were structurally optimised before QED corrections were included perturbatively.
Equilibrium bond lengths obtained are reported in the table.
While VP and SE show growing contributions with increasing nuclear charge $Z$ of
the metal atom, their sign is not always opposite to each other. In the case of
BeF, VP and SE both show a negative sign. This is also the only case,
in which the metal has a smaller nuclear charge than the fluoride. Besides that,
the results are in accordance with general trends expected based on QED corrections for the
corresponding group 2 atoms and agree also within 15~\% with the very recently reported specific case 
of RaF for which a total QED correction of $-6.4626$~meV was obtained on the four-component Fock-space
coupled cluster level \cite{wilkins:2024}.

\subsection{Electronic transition energies of BaF and RaF}
The QED contributions to the $\Pi_{1/2}\leftarrow \Sigma_{1/2}$ and
$\Pi_{3/2}\leftarrow \Sigma_{1/2}$ transition energies of BaF and RaF are
listed in Table \ref{bafraf}. In the case of BaF, 216.00 pm and 218.00 pm
were chosen as bond length for the ground and excited state, respectively, of the molecule.
For RaF, 223.84 pm was used for both states. The distances were chosen 
to provide a better comparison to the calculations of Skripnikov, Chubukov and 
Shakhova \cite{leonid:2021, skripnikov:2021, zaitsevskii:2022}, who used them as well.
The obtained QED contributions to the BaF transitions differ from the ones
calculated by Skripnikov \cite{skripnikov:2021} by $\approx$ 23\% for the
$\Pi_{1/2}\leftarrow \Sigma_{1/2}$ and $\approx$ 24\% for the
$\Pi_{3/2}\leftarrow \Sigma_{1/2}$ transitions. The QED
contributions to the transition energies of RaF
differ from the ones calculated by Zaitsevskii \textit{et al.} \cite{zaitsevskii:2022} 
by $\approx$ 16\% for the $\Pi_{1/2}\leftarrow \Sigma_{1/2}$ and $\approx$ 19\%
for the $\Pi_{3/2}\leftarrow \Sigma_{1/2}$ transitions.
The larger deviations between the values are attributed to the different
formulation of the wave function and the SE. The latter's share of the QED correction is larger than the
one of the VP.\\

\section{Conclusion}

The Uehling potential, the Pyykkö--Zhao and the Flambaum--Ginges effective potentials
have been implemented successfully in a two-component ZORA framework. For the
Flambaum--Ginges potential, repeated numerical evaluation of 
integrals in the magnetic and high-frequency contributions, for which no analytical solution is available, was evaded by using
cubic spline interpolation instead. Combined with the ZORA framework, this allows
for fast evaluation of said QED corrections in atoms and molecules. The various
exemplary calculations demonstrate the accuracy of this approach. Corrections
to energy levels, ionisation energies and transition energies compare well
to four-component numerical DHF calculations by Thierfelder and Schwerdtfeger \cite{thierfelder:2010}
or four-component AOC-HF calculations by Sunaga, Salman and Saue \cite{sunaga:2022}.
Here, it was observed, that the basis set had a greater impact on the QED corrections,
than the level of theory alone. 
Furthermore, the contributions to the ionisation energies of 
group 2 monofluorides are presented.
Overall, we find the impact of different schemes to account for electron self-energy corrections to be much larger than
residual deviations between our present two-component ZORA framework and related four-component frameworks. 
The two-component scheme reported herein thus opens up an avenue to assess the role of QED corrections in atoms
and larger molecules as well as to explore applicabilities and limitations of various approximate schemes to account for 
contributions from quantum electrodynamics to energies and molecular properties.

\section{Acknowledgement}
The authors thank G. A. Aucar, I. A. Aucar and K. Kozio\l{} for discussions and gratefully acknowledge
K. Kozio\l{} as well as A. Kiuberis for sharing results with increased number of digits in personal communications. P. S. thanks
Prof. P. Indelicato (Laboratoire Kastler Brossel, Sorbonne, Paris) for a visiting professorship.

\bibliographystyle{apsrev4-2}

%
\end{document}